\documentclass[envcountsect,11pt,letterpaper]{llncs}
\usepackage{soul}
\usepackage[utf8]{inputenc}
\usepackage[table]{xcolor}
 \usepackage{cancel} 
\usepackage{float}
\usepackage[margin=1in]{geometry}
\usepackage{graphicx}
\usepackage{xcolor}
\usepackage{amsmath}
\usepackage{amssymb}
\usepackage[sort,nocompress]{cite}
\usepackage[ruled,vlined,lined,boxed,commentsnumbered]{algorithm2e}
\usepackage{todonotes}
\usepackage{verbatim}
\usepackage{subcaption}
\usepackage[normalem]{ulem}
\usepackage{array}
\usepackage{algorithmic}
\usepackage{mathrsfs}
 \usepackage{multirow}
 \usepackage{blindtext}
\usepackage{hyperref}
\usepackage{graphicx}
\usepackage{todonotes}

\usepackage{pgf}
\usepackage{tikz}
\usepackage{tikz-cd}
\tikzcdset{scale cd/.style={every label/.append style={scale=#1},
    cells={nodes={scale=#1}}}}

\usetikzlibrary{arrows,automata,positioning}
\SetKwComment{Comment}{$\triangleright$\ }{}

\newcommand{\LCP}{\mathsf{LCP}}
\newcommand{\lcp}{\mathop{\mathsf{lcp}}}

\title{Enhanced Graph Pattern Matching}

\author{
Nicola Cotumaccio
} 

\institute{
}

\date{\today}

\begin{document}

\maketitle
\thispagestyle{empty}

\begin{abstract}
Pattern matching queries on strings can be solved in linear time by Knuth–Morris–Pratt (KMP) algorithm. In 1973, Weiner introduced the suffix tree of a string [FOCS 1973] and showed that the seemingly more difficult problem of computing matching statistics can also be solved in liner time. Pattern matching queries on graphs are inherently more difficult: under the Orthogonal Vector hypothesis, the graph pattern matching problem cannot be solved in subquadratic time [TALG 2023]. The complexity of graph pattern matching can be parameterized by the topological complexity of the considered graph, which is captured by a parameter $ p $ [JACM 2023].

In this paper, we show that, as in the string setting, computing matching statistics on graph is as difficult as solving standard pattern matching queries. To this end, we introduce a notion of longest common prefix (LCP) array for arbitrary graphs.
\end{abstract}

\setcounter{page}{1}

\section{Introduction}

The increasing amount of data has motivated a considerable effort to improve the available string processing algorithms (and in particular, the available string searching algorithm). Pattern matching is one of the fundamental problems to solve on a text, and it can be classically solved in linear time by using Knuth–Morris–Pratt (KMP) algorithm \cite{knuthscicomp1977}. Knuth-Morris-Pratt algorithm was only the start point of a terrific investigation of the problem. We can identify the following lines of research.
\begin{itemize}
    \item \emph{Enhanced pattern matching}. From the beginning of the seventies mark, there was a growing effort to understand whether it is possible to efficiently solve more complicated variants of the pattern matching problem, contributing to the development of a variety of elegant algorithmic techniques (see \cite{crochemorebook2021} for a recent book on the topic). The breakthrough result occurred in 1973, when Weiner invented \emph{suffix trees}, showing that the suffix trees support a linear-time solution of the problem (today known as the problem) of computing matching statistics. Over the years, suffix trees proved able to solve a variety of problems, including computing longest common extensions, maximal repeats, minimal absent words, maximal palindromes and the LZ77 factorization of a string (see \cite{gusfieldbook1997, makinenbook2023, navarrobook2016, apostolicocaw1985}). Some variants of suffix trees support more difficult problems, such as parameterized pattern matching \cite{bakerstoc1993}. Nonetheless, suffix trees are not suitable to handle some other natural variations, such as the string matching with don't care problem, where both the pattern and the text may contain wildcards. The classical solution of Fischer and Paterson, based on fast Fourier transform, runs in time $ O(n \log w \log \sigma) $, where $ n $ is the length of the text, $ w $ is the length of the pattern, and $ \sigma $ is the size of alphabet \cite{fischermit1974}. Indyk removed the dependence on the alphabet by giving a randomized algorithm $ O(n \log n) $ algorithm \cite{indykfocs1998}; then, Kalai gave a randomized $ O(n \log w) $ algorithm \cite{kalaisoda2002}. Finally, Cole and Hariharan gave a deterministic $ O(n \log w) $ algorithm \cite{colestoc2002, colescicomp2003}, later simplified in \cite{cliffordipl2007}.

    \item \emph{Graph pattern matching} The classic formulation of the pattern matching problem can be seen as a special instance of the problem of determining whether a string can be read by following edges in a node-labeled or edge-labeled graph, which in turn is strictly connected to the problem of deciding whether a string is accepted by a finite automaton --- in the string setting, the text string can be interpreted as a graph only consisting of a single path. The graph pattern matching problem was initially motivated by the need of retrieve information stored in hypertext \cite{manber1992, tasuya1993, park1995, amir2000, rautiainen2017, navarro2000}. Assuming a constant alphabet, pattern matching on labeled trees can be solved in linear time \cite{tasuya1993}. Amir et al. considered the general setting in which nodes can be labeled with nonempty strings, showing that pattern matching queries on arbitrary graphs can be solved in $ O(\mathfrak{e} + |\pi| e) $ time, where $ e $ is the number of edges in the graph, $ |\pi| $ is the length of the pattern, and $ \mathfrak{e} $ is the total length of all labels in the graph \cite{amir2000}.

    A related extension is the \emph{tree pattern matching problem}. In the literature, the three pattern matching problem is not the problem of determining whether a string occurs in a tree, but it is the problem of deciding whether a labeled tree of size $ w $ occurs as a subtree of an another labeled tree of size $ n $. The tree pattern matching problem finds a variety of applications, including symbolic computation, code optimization of compilers and implementations of abstract data types \cite{hoffmanjacm1982}. The naive algorithm runs in $ O(wn) $ time. Kosaraju designed an $ \tilde{O}(w^{0.75}n) $ algorithm \cite{kosarajufocs1989}; Dubiner et al. gave an $ \tilde{O}(\sqrt{w} n) $ algorithm \cite{dubinerjacm1994, dubinerfocs1990}; Cole and Hariharan exhibited an $ O(n \log w) $ algorithm through a reduction to the subset matching problem \cite{colestoc2002, colescicomp2003} (see also \cite{indykfocs1997, indykfocs1998}). The tree matching problem stimulated many new algorithmic techniques; for example, in his paper, Kosaraju introduced the suffix tree of a tree and the convolution of a tree and a string \cite{kosarajufocs1989}.
    \item \emph{Compressed pattern matching} The field of compressed data structures aims to compress data in such a way that it is possible to retrieve information without needing to decompress the stored data \cite{navarrobook2016, makinenbook2023, adjerohbook2008}. Being able to solve pattern matching on \emph{compressed} data is arguably of the most important queries, and possibly the most studied problem in the compressed data realm. Storing the suffix tree of a string requires $ O(n \log n) $ bits (where $ n $ is length of the string), and the hidden constant in the space bound is generally too big for big data applications. Manber and Myers introduced \emph{suffix arrays}, which do not have the full functionality of suffix trees, but only require $ n \log n $ bits while still supporting conventional pattern matching queries \cite{manberscicomp1993, manbersoda1990}. The year 2000 was a crucial one for compressed pattern matching: on the one hand, Grossi and Vitter showed how to compress the text and the suffix array while supporting string searching queries \cite{grossistoc2000, grossiscicomp2005}; on the other hand, Ferragina and Manzini invented the FM-index \cite{ferraginafocs2000, ferraginajacm2005}, which also supports string searching queries on a compressed representation based on the Burrows-Wheeler Transform \cite{burrows1994}. Subsequently, Sadakane showed how to compress \emph{suffix trees} while retaining their full functionality \cite{sadakanetcsyst2007, sadakanesoda2002}, notably improving on previous work \cite{munrojalg2001, grossiscicomp2005}. Subsequently, a plethora of (slightly different) compressed suffix trees were proposed, with different theoretical and practical space-time tradeoffs (for example, \cite{fischercpm2008, fischertcs2009, fischeripl2010, russotalg2011, russolata2008, gogexp2013, abeliukalgorithms2013, navarrodcc2014}), which also depend on the specific variant of compressed suffix array used in the compressed suffix tree (see \cite{navarrosurveys2007}). Recently, Gagie et al. showed how to design a fully-functional suffix tree with space and time bound parameterized by the number $ r $ of runs in the Burrows-Wheeler transform \cite{gagie2020jacm, gagiesoda2018}.
\end{itemize}

Distinct variants of the pattern matching problem have distinct complexities and require notably different approaches, which suggests that these problems can represent a benchmark to test the limits of algorithmic techniques. When it comes to lower bounds, a recent and fruitful line of research has shown that the complexity of string processing problems can be often determined assuming (some variant of) the Strong Exponential Time Hypothesis or the Orthogonal Vector Hypothesis. Under these conjectures, multiple problems cannot be solved in strongly subquadratic time, including computing local alignments \cite{abboudicalp2014}, the Fréchet distance \cite{bringmannfocs2014}, the edit distance \cite{backursscicomp2018, backursstoc2015}, the dynamic time warping distance and longest common subsequences \cite{abboudfocs2015}, longest palindromic subsequences and longest tandem subsequences \cite{bringmannfocs2015}. These techniques were adapted to prove the conditional hardness of several other problems, including longest common increasing subsequences \cite{durajalgorithmica2019, durajipec2017}, elastic-degenerate text \cite{gibneyspire2020}, episode matching \cite{billecpm2022}, Wheeler languages \cite{beckerspire2023} (see \cite{patrascustoc2010, amirfocs2014, abboudstocs2016, backursfocs2016, polakipl2018, kociumakaalgorithmica2019} for more conditional results with the same flavor).

When it comes to lower bounds for pattern matching problems (conditional to the Strong Exponential Time Hypothesis or the Orthogonal Vector Hypothesis), once again small variation of the problem can lead to drastic differences. For example, while the tree pattern matching problem can be solved in quasilinear time (as we saw before), if we consider the unlabeled case --- that is, if we consider the subtree isomorphism problem --- then the problem cannot be solved in subquadratic time \cite{abboudtalg2018, abboudsoda2016}. Recently, Equi et al. showed that the graph pattern matching problem cannot be solved in subquadratic time \cite{equitalg2023, equiicalp2019, equitcs2023, equisofsem2021}.

\section{Our results}

In view of Equi et al. result, the pattern matching problem on graphs is inherently more difficult that the pattern matching on strings. Cotumaccio et al. introduced a parameter $ p $ associated to a graph that refines the complexity of pattern matching --- between linear and quadratic, depending on $ p $ \cite{cotumacciojacm2023, cotumacciosoda2021}. The natural question is whether more complex variant problem of graph pattern matching problem are still tractable or not.

Let us consider the problem originally considered by Weiner when introducing suffix trees: computing matching statistics. Weiner showed that, even though the problem of computing matching statistics appears more difficult that conventional pattern matching, it can still be solved in linear time by using a suffix tree.

We can prove that, even on graphs, matching statistics can be computed as efficiently as solving conventional pattern matching queries. Here we will only provide the main intuition needed to prove our claim; the full proof is quite technical, and will be provided in the extended version of this paper.

In the string setting, matching statistics can be computed using the Burrows-Wheeler Transform and the LCP array \cite{ohlebuschspire2010}. Since Cotumaccio et al.'s work generalizes the Burrows-Wheeler Transform to graphs, our first step is to generalize the LCP array to graphs. Consider a graph $ G $ with $ n $ nodes, and assume without loss of generality that  every node of $ G $ has at least one incoming edge. For every node $ u $, we consider the lexicographically smallest string $ \min_u $ in $ \Sigma^\omega $ that can be read starting from $ u $ and following edges in a backward fashion. We sort all the strings $ \min_u $'s, obtain strings $ \alpha_1 \preceq \alpha_2 \preceq \dots \preceq \alpha_n $. Then, we define the array $ \LCP^{\min}_G[2, n] $ such that $ \LCP^{\min}_G[i] = \lcp(\alpha_{i - 1}, \alpha_i) $. Analogously, we define an array $ \LCP^{\max}_G[2, n] $ by considering the lexicographically \emph{largest} string reaching each node. Note that it would be possible to store more information about $ G $ by defining a \emph{unique} LCP array of size $ 2n - 1 $ obtained by jointly sorting the minimum and the maximum string reaching each node.
 
In the graph setting, however, the LCP array is not sufficient, because each node is potentially reached by infinitely many paths. The crux of our proof is to show that it is possible to mantain $ O(p) $ values during the execution of our algorithm in such a way that (i) we can compute matching statistics and (ii) the $ O(p) $ values can be quickly updated.

In the full version we will prove the following theorem.

\begin{theorem}
    Let $ G $ be a graph with parameter $ 1 \le p \le n $ \cite{cotumacciojacm2023}, where $ n $ is the number of vertices. Then, there exists a data structure such that, given a string of length $ w $, we can compute the matching statistics of $ w $ with respect to $ G $ in $ O(wp^2 \log \log (p \sigma)) $ time, where $ \sigma $ is the size of the alphabet.
\end{theorem}

\bibliographystyle{plainurl}
\bibliography{main.bib}

\end{document}